\documentclass[
 reprint,
 amsmath,
 amssymb,
 superscriptaddress,
 aps,
 groupedaddress
]{revtex4-1}

\usepackage[utf8]{inputenc}
\usepackage{graphicx}
\usepackage{bm}
\usepackage{gensymb}
\usepackage{float}

\usepackage{array,mathtools,amssymb,booktabs}
\newcolumntype{C}{>{$}c<{$}}
\AtBeginDocument{
\heavyrulewidth=.08em
\lightrulewidth=.05em
\cmidrulewidth=.03em
\belowrulesep=.65ex
\belowbottomsep=0pt
\aboverulesep=.4ex
\abovetopsep=0pt
\cmidrulesep=\doublerulesep
\cmidrulekern=.5em
\defaultaddspace=.5em
}
\usepackage{bm}
\usepackage{float}
\usepackage{hyperref}
\usepackage{siunitx}
\usepackage{threeparttable}
\begin{document}

%\preprint{APS/123-QED}\

\title{Direct measurement of the $7s\ ^2S_{1/2}\rightarrow$ $7p\ ^2P_{3/2}$ transition frequency in $^{226}$Ra$^{+}$}
\author{C. A. Holliman}
\email{cholliman3@gmail.com}
\author{M. Fan}
\author{A. Contractor}
\author{M. W. Straus}
\author{A. M. Jayich}
\affiliation{Department of Physics, University of California, Santa Barbara, Santa Barbara, California 93106, USA}
\affiliation{California Institute for Quantum Entanglement, Santa Barbara, California 93106, USA}

\begin{abstract}
    We report a direct measurement of the $7s\ ^2S_{1/2}\rightarrow$ $7p\ ^2P_{3/2}$ electric dipole transition frequency in $^{226}$Ra$^{+}$. With a single laser-cooled radium ion we determine the transition frequency to be \SI{785722.11\pm0.03}{\giga\hertz} by directly driving the transition with frequency-doubled light and measuring the frequency of the undoubled light with an iodine reference. This measurement addresses a discrepancy of five combined standard deviations between previously reported values.
\end{abstract}

\date{\today}

\maketitle

\begin{figure}[h]
    \centering
    \includegraphics[width=0.85\linewidth]{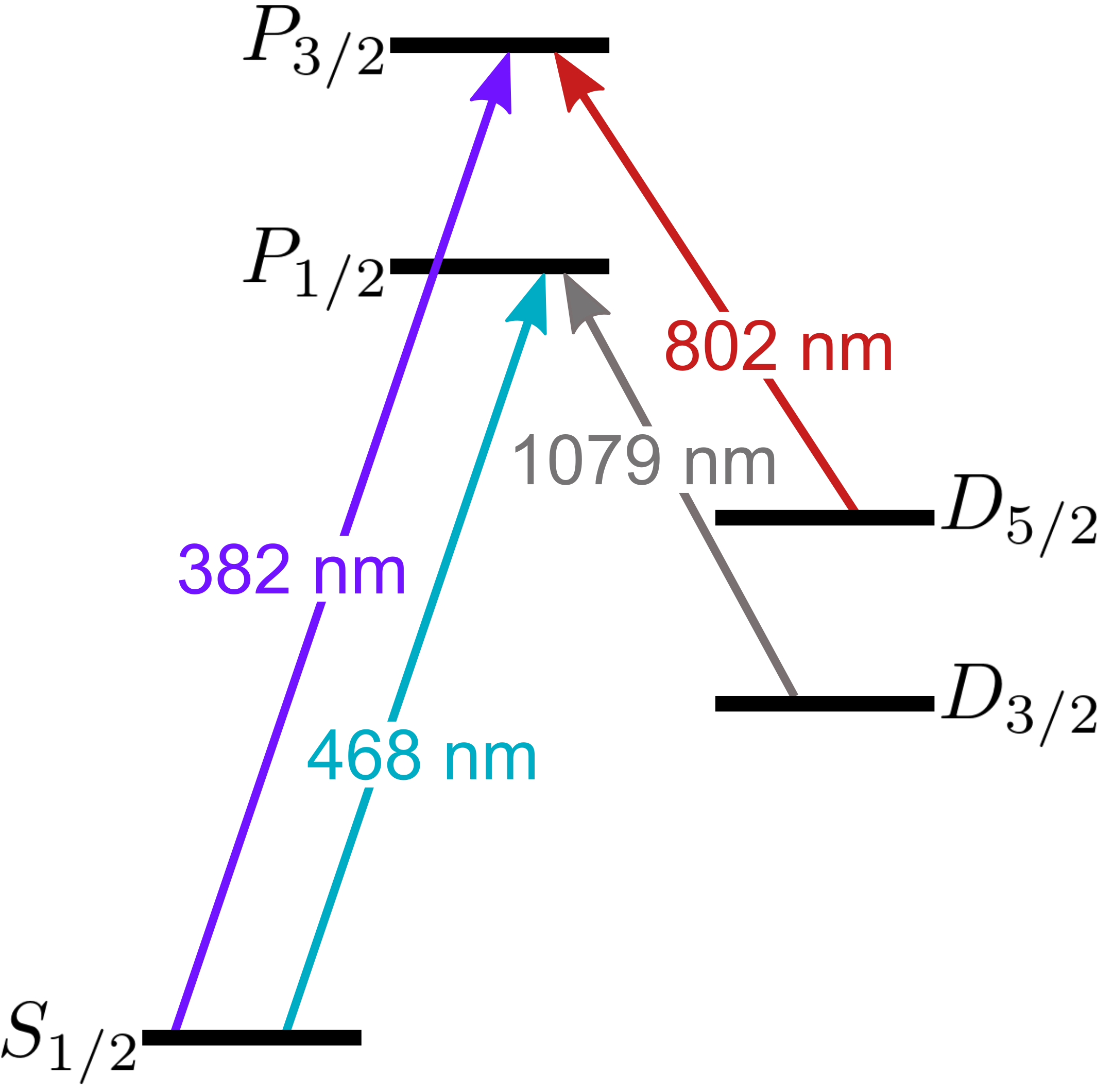}
    \caption{The Ra$^{+}$ energy-level structure with the transitions driven in the measurement of the $7s\ ^2S_{1/2}\rightarrow$ $7p\ ^2P_{3/2}$ transition frequency.}  
    \label{fig:energy_level}
\end{figure}

Trapped radium ions are appealing for quantum information science \cite{Hucul2017}, searches for new physics beyond the standard model \cite{Fortson1993, Berengut2018}, and precision timekeeping \cite{Fan2019, Sahoo2009a, NunezPortela2014}.  Ra$^{+}$ is a potential candidate for a transportable optical clock, as the transitions can all be addressed with direct diode lasers, and the ion's laser cooling and fluorescence transition at 468 nm is far from the UV compared to most trapped ion optical clock candidates \cite{Ludlow2015}, see Fig. \ref{fig:energy_level}.  The narrow electric quadrupole transitions, the $7s\ ^2S_{1/2}\rightarrow$ $6d\ ^2D_{5/2}$ transition at 728 nm and the $7s\ ^2S_{1/2}\rightarrow$ $6d\ ^2D_{3/2}$ transition at 828 nm, may be used to study radium's nuclear structure \cite{Heilig1974, Reinhard2020} or set bounds on sources of new physics \cite{Berengut2018}. 

There is a discrepancy between two previously reported values of the $7s\ ^2S_{1/2}\rightarrow$ $7p\ ^2P_{3/2}$ (382-nm) transition frequency in $^{226}$Ra$^+$, \SI{785722.10\pm0.03}{\giga\hertz} \cite{Holliman2019} and \SI{785721.670\pm0.070}{\giga\hertz} \cite{NunezPortela2014}, which is included in a compilation of radium spectroscopy data \cite{Dammalapati2016}.  The former value was determined from a sum of two direct frequency measurements in $^{226}$Ra$^+$ of the $S_{1/2}\rightarrow D_{5/2}$ and $D_{5/2}\rightarrow P_{3/2}$ transitions. The latter value was determined using the transition frequency calculated from a sum of direct measurements in $^{214}$Ra$^+$ \cite{NunezPortela2014} and the isotope shift between $^{214}$Ra$^+$  and $^{226}$Ra$^+$ measured by Neu, \textit{et al.} \cite{Neu1988}.  In this work we help resolve this discrepancy with direct spectroscopy at 382 nm of the $S_{1/2}\rightarrow P_{3/2}$ transition with a single laser-cooled radium ion.  The frequency measurement is calibrated by absorption spectroscopy with known transitions in molecular iodine.  The value measured in this work, \SI{785722.11\pm0.03}{\giga\hertz}, agrees with the value from the sum of direct measurements in $^{226}$Ra$^+$ \cite{Holliman2019}. 

\begin{figure}[h]
    \centering
    \includegraphics[width=0.84\linewidth]{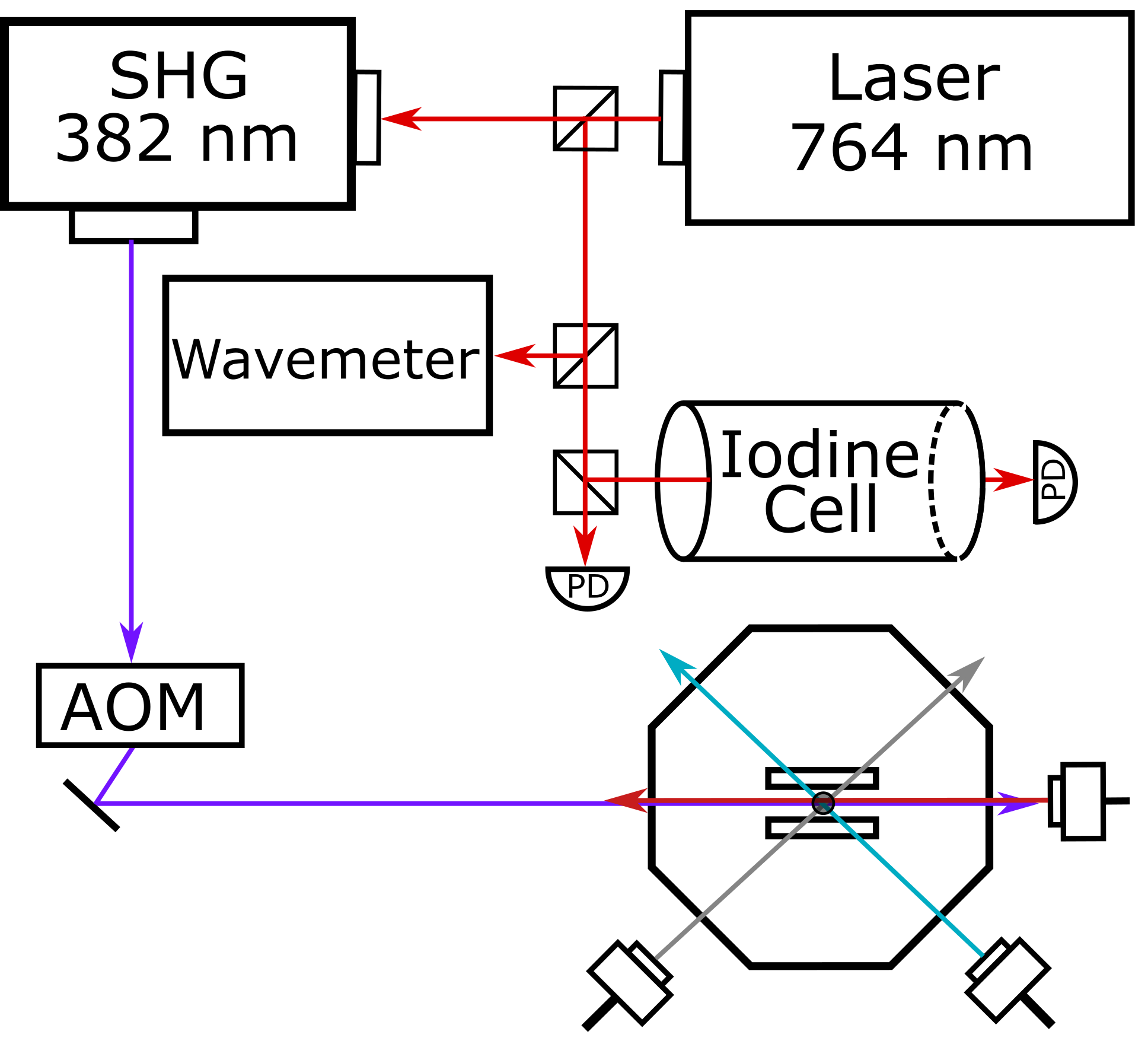}
    \caption{Diagram of the measurement setup.  Light at 764 nm (red) from a  Ti:sapphire laser is frequency doubled via second harmonic generation (SHG) to drive the  $S_{1/2}\rightarrow P_{3/2}$ transition in Ra$^{+}$ at 382 nm (purple).  An acousto-optic modulator (AOM) is used as a shutter for the 382-nm spectroscopy light.  The 764-nm frequency is recorded on a wavemeter.  A photodiode (PD) before the iodine cell compensates for laser power drifts. Other beams addressing the ions: Doppler cooling (teal), the $D_{3/2}\rightarrow P_{1/2}$ repump (gray), and the $D_{5/2}\rightarrow P_{3/2}$ cleanout (dark red).}  
    \label{fig:setup}
\end{figure}

\begin{figure}[h]
    \centering
    \includegraphics[width=1.\linewidth]{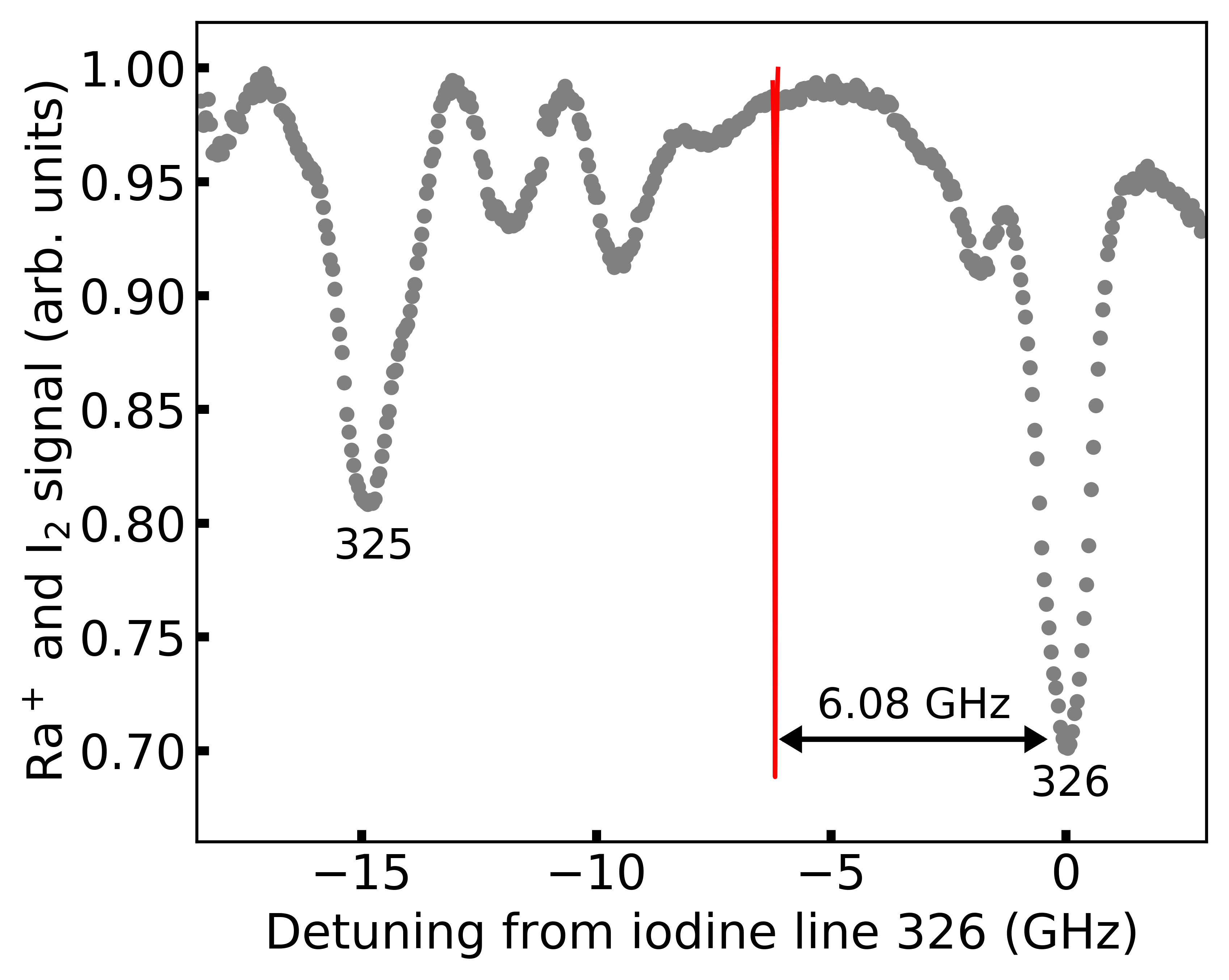}
    \caption{Iodine absorption (gray) and radium spectroscopy (red) data plotted in terms of the undoubled 764-nm laser frequency that we record on a wavemeter. Data are scaled and vertically offset for clarity. The iodine lines closest to the transition are lines 325 and 326 \cite{Gerstenkorn1982}.}  
    \label{fig:joint}
\end{figure}

The iodine linear absorption spectrum is measured around 764 nm with light from a Ti:sapphire laser.  Frequency-doubled light from the same laser drives the $S_{1/2}\rightarrow P_{3/2}$ transition of $^{226}$Ra$^+$ at 382 nm (see Fig. \ref{fig:setup}). From iodine reference lines near 764 nm we can determine the $S_{1/2}\rightarrow$ $P_{3/2}$ transition frequency.  During both the radium and iodine spectroscopy the 764-nm laser's frequency is recorded with a wavemeter (High Finesse WS-8), see Fig. \ref{fig:joint}.

To measure the $S_{1/2}\rightarrow$ $P_{3/2}$ transition frequency we first scan the iodine spectrum near 764 nm, at half of the transition frequency  reported in \cite{Holliman2019}. We scan over a 20-GHz range that includes two iodine reference lines, 325 and 326 \cite{Gerstenkorn1982}.  This is followed by radium spectroscopy and then a second iodine scan to account for wavemeter drift.  The radium spectroscopy uses state detection to determine if the population is in a ``bright'' state, $S_{1/2}$ or $D_{3/2}$, or in the $D_{5/2}$ ``dark''  state, by collecting Ra$^{+}$ fluorescence at 468 nm onto a photomultiplier tube.  The $S_{1/2}\rightarrow$ $P_{3/2}$ transition frequency is calculated from the difference between the radium 382-nm line center and twice the frequency of iodine line 326.

\begin{figure}[h]
    \centering
    \includegraphics[width=1.\linewidth]{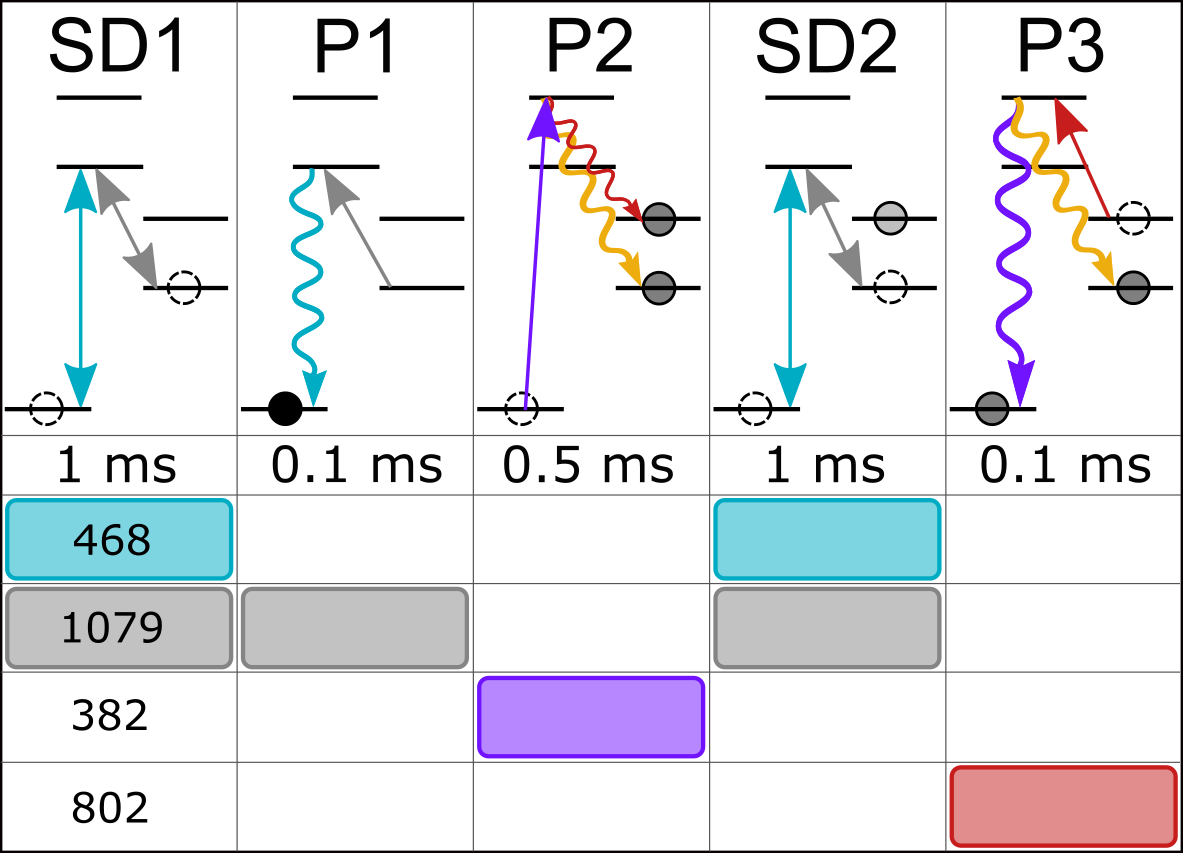}
    \caption{Pulse sequence used for the $S_{1/2}\rightarrow P_{3/2}$ spectroscopy. Squiggly lines depict electric dipole allowed decays, straight lines with an arrow indicate optical pumping, and double arrows indicate optical cycling. The optical pumping (P) and state detection (SD) are described in the text.}  
    \label{fig:ps382}
\end{figure}

\begin{figure}[h]
    \centering
    \includegraphics[width=1.\linewidth]{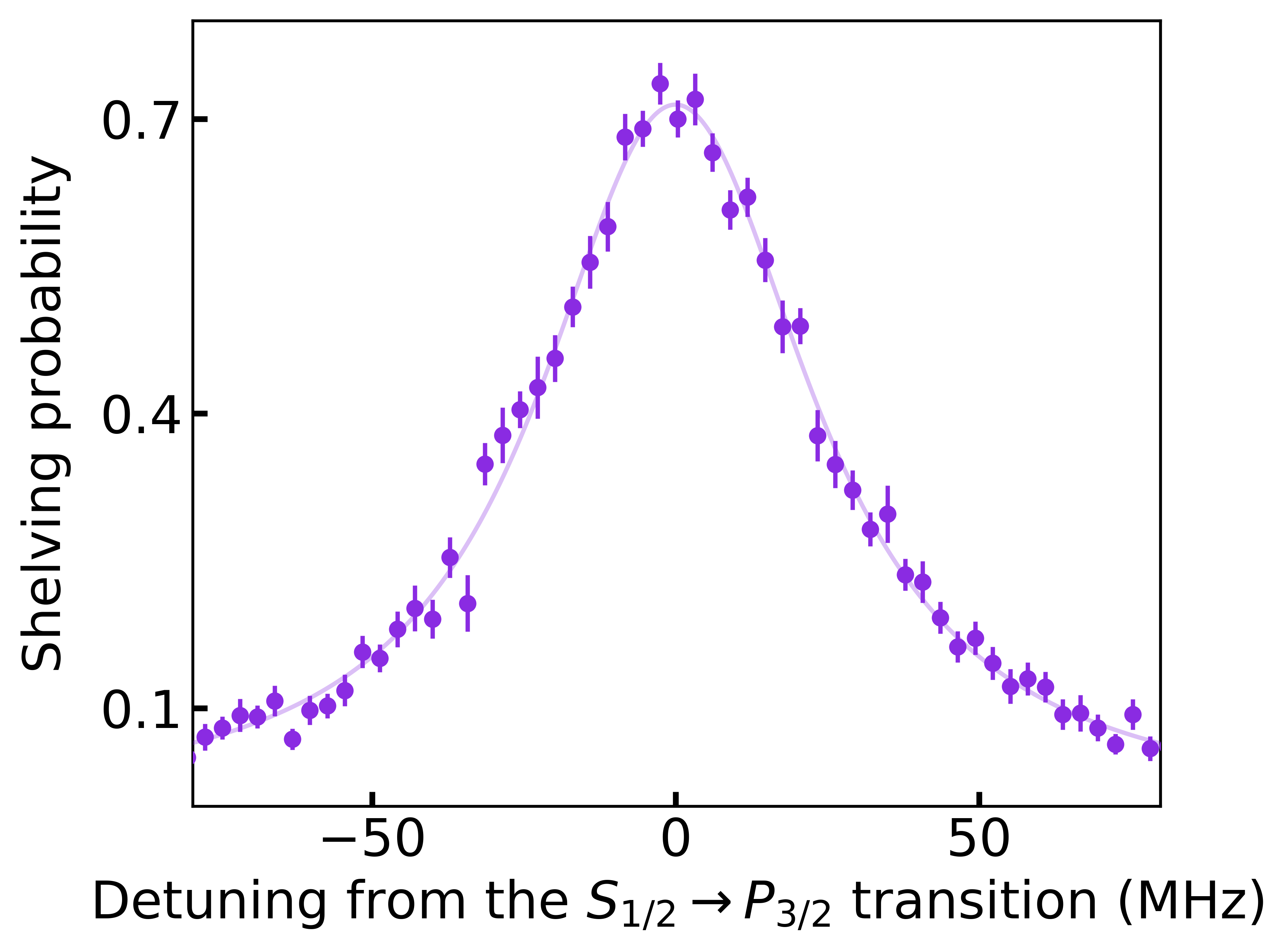}
    \caption{Spectroscopy of the $S_{1/2}\rightarrow P_{3/2}$ transition at 382 nm. The Lorentzian fit that accounts for $P_{3/2}$ state depletion gives a FWHM of \SI{40\pm1}{\mega\hertz} and a reduced $\chi^2$ of 1.06. Error bars are the most likely 68\% confidence interval of a binomial distribution. The shelving probability is limited by branching fractions from the $P_{3/2}$ state, as well as the probe duration, which was chosen to minimize power broadening while still realizing a reasonable signal-to-noise ratio.}  
    \label{fig:spectrum}
\end{figure}

The pulse sequence for the $S_{1/2}\rightarrow P_{3/2}$ frequency measurement is shown in Fig. \ref{fig:ps382}.  All light is linearly polarized to drive symmetric Zeeman transitions with the same amplitude, with a 3-G magnetic field along the trap's axial direction.  We Doppler cool the ion for \SI{500}{\micro\second} before each pulse sequence.  The initial state detection determines if the ion is cooled and if the population is in a bright state (SD1). If the population is not initialized in a bright state or if the ion is in a large orbit the data point is excluded.  Population in the $D_{3/2}$ state is then optically pumped for \SI{100}{\micro\second} with light at 1079 nm to the ground state (P1).  The $S_{1/2}\rightarrow P_{3/2}$ spectroscopy transition is then driven with light at 382 nm for \SI{500}{\micro\second} (P2).  Decays from the $P_{3/2}$ state have an 11\% probability to be shelved in the $D_{5/2}$ dark state \cite{Fan2019a}.  We use these decays to determine the probability of driving the $S_{1/2}\rightarrow$ $P_{3/2}$ transition at a given laser frequency with a second state detection (SD2) to form a binomial distribution, see Fig. \ref{fig:spectrum}.  Finally, any population that remains in the $D_{5/2}$ state is reset to the bright states by optically pumping with 802 nm light (P3).  The pulse sequence is repeated \SI{32000}{} times as the 382-nm laser is swept over the transition in a $\sim$\SI{100}{\mega\hertz} range.  The $S_{1/2}\rightarrow P_{3/2}$ spectroscopy is shown in Fig. \ref{fig:joint}, along with the corresponding iodine absorption reference spectrum. 

The $S_{1/2}\rightarrow P_{3/2}$ radium measurement data are binned by frequency, and the shelving probability is fit to the exponential of a Lorentzian, see Fig. \ref{fig:spectrum}, to account for population depletion of the $S_{1/2}$ state (see Appendix A). The full width at half maximum of \SI{40\pm1}{\mega\hertz} gives a lower bound of \SI{4.0\pm0.1}{\nano\second} for the $P_{3/2}$ state lifetime, which is in agreement with the calculated value of \SI{4.73}{\nano\second} \cite{Pal2009}.  The 382-nm light is incident on the ion along the trap's axial direction (see Fig. \ref{fig:setup}) to minimize micromotion broadening of the transition \cite{Berkeland1998}.

The closest iodine reference line to the $^{226}$Ra$^{+}$ $S_{1/2}\rightarrow P_{3/2}$ transition is line 326 \cite{Gerstenkorn1982}. We calibrate the 326 line center frequency with a wavemeter to an absolute frequency using IodineSpec5 \cite{IodineSpec5, Knoeckel2004}. We fit the absorption dip of line 326 in the IodineSpec5 data to a Voigt function to determine its frequency-doubled line center, \SI{785734.265\pm0.003}{\giga\hertz}.

The frequency difference between the $S_{1/2}\rightarrow P_{3/2}$  line center at 382 nm and twice the value of the iodine line 326 center frequency is \SI{12.16\pm0.03}{\giga\hertz} (see Fig. \ref{fig:joint}). From this frequency difference we determine the $S_{1/2}\rightarrow P_{3/2}$ transition frequency to be \SI{785722.11\pm0.03}{\giga\hertz}. The uncertainty is $\sigma_{\text{total}} = \sqrt{\sigma_{\text{Ra}^{+}}^{2} + \sigma_{\text{I}_{2}}^2 + \sigma_{\text{spec}}^{2} + \sigma_{\text{wm}}^{2}}$,  where $\sigma_{\text{Ra}^{+}}$ (20 MHz) is half the full width at half maximum of the fitted Ra$^{+}$ spectrum (see Fig. \ref{fig:spectrum}) to account for possible magnetic field shifts due to Zeeman levels, $\sigma_{\text{I}_{2}}$ (10 MHz) is the iodine spectroscopy fitting uncertainty, $\sigma_{\text{spec}}$ (3 MHz) is the IodineSpec5 line uncertainty, and $\sigma_{\text{wm}}$ (20 MHz) is the wavemeter uncertainty.

\begin{table}[h] \centering
\begin{threeparttable}
\caption{Summary of reported $S_{1/2}\rightarrow P_{3/2}$ transition frequencies in $^{226}$Ra$^{+}$. All values are offset from \SI{785722000}{\mega\hertz}.}
\label{table:382_frequencies}
\begin{ruledtabular}
\begin{tabular}{lcccc}
%\rowcolor{gray!50}
    Transition & \cite{Rasmussen1933} & \cite{NunezPortela2014} & \cite{Holliman2019} & This work \\ 
    \midrule
    $S_{1/2}\rightarrow P_{3/2}$ & \SI{1000\pm4000}{} & \SI{-330\pm70}{}\tnote{a} & \SI{100\pm30}{}\tnote{a} & \SI{110\pm30}{} \\
\end{tabular}
\end{ruledtabular}
\begin{tablenotes}
\item [a]Frequencies calculated from indirect measurements
\end{tablenotes}
\end{threeparttable}
\end{table}

This work measures a transition that was previously directly measured nearly a century ago \cite{Rasmussen1933} and addresses a discrepancy between the frequencies reported by Nu\~nez, \textit{et al.} \cite{NunezPortela2014} and by Holliman, \textit{et al.} \cite{Holliman2019}, see Table \ref{table:382_frequencies}.  This $^{226}$Ra$^{+}$ frequency measurement also serves as a check for the $S_{1/2}\rightarrow D_{5/2}$ (728 nm), $S_{1/2}\rightarrow D_{3/2}$ (828 nm), $D_{3/2}\rightarrow P_{3/2}$ (708 nm), and $D_{5/2}\rightarrow P_{3/2}$ (802 nm) transition frequencies measured in \cite{Holliman2019}. The $S_{1/2}\rightarrow P_{3/2}$ transition frequency can be calculated from the sum of the 728- and 802-nm frequencies, \SI{785722.10\pm0.03}{\giga\hertz}, or the 828- and 708-nm frequencies, \SI{785722.07\pm0.05}{\giga\hertz}. Both values derived from \cite{Holliman2019} agree with the reported direct measurement of the $S_{1/2}\rightarrow P_{3/2}$ transition frequency.  We verify the self-consistency of these sets of measurements in Appendix B.

We acknowledge use of the UC Santa Barbara physics department's Broadly-tunable Illumination Facility for Research, Outreach, Scholarship, and Training Ti:sapphire laser, which provided the 764- and 382-nm light used in this work. We also acknowledge support from NSF Grant No. PHY-1912665 and the UC Office of the President (Grant No. MRP-19-601445).

\section*{Appendix A: Depletion Analysis}\label{appendix:dep}
If a large proportion of the population is driven out of the $S_{1/2}$ state during driving to the $P_{3/2}$ state, the expected Lorentzian line shape of the spectroscopy distorts due to decays from the $P_{3/2}$ state to the metastable $D_{3/2}$ and $D_{5/2}$ states. If we consider the $S_{1/2}$ and $P_{3/2}$ states as a two-level system and drive the $S_{1/2} \rightarrow P_{3/2}$ transition for a time comparable to the $P_{3/2}$ state lifetime or longer so that the system approaches equilibrium, there is a fraction of the population that remains in the excited state, $\text{R}_{P}$. When we take into account decays to the metastable $D_{3/2}$ and $D_{5/2}$ states, the $P_{3/2}$ state population $\text{P}_{P}$ is given by
\begin{equation}
\begin{aligned}
\text{P}_{P}= \text{R}_{P}(1-\text{P}_{D}),\\
\end{aligned}
\end{equation}
where $\text{P}_{D}$ is the total population in both the $D_{3/2}$ and $D_{5/2}$ states. We can approximate the decay into the $D$ states as
\begin{equation}
\begin{aligned}
    \frac{d\text{P}_{D}}{dt}=&\frac{\text{P}_{P}(1-r)}{\tau}\\
    =&\frac{\text{R}_{P}(1-\text{P}_{D})(1-r)}{\tau},
\end{aligned}
\end{equation}
where $\tau$ is the lifetime of the $P_{3/2}$ state and $r$ is the branching fraction of the $P_{3/2}$ state to the $S_{1/2}$ state. Under the assumption that the population is always initialized to the $S_{1/2}$ state this gives
\begin{equation}
    \text{P}_{D} = 1-\text{exp}\left(\frac{-t\text{R}_{P}(1-r)}{\tau}\right).
\end{equation}
The final dark state population $\text{P}_{D_{5/2}}$ is determined from the branching fractions of the $P_{3/2}$ state to the $D_{5/2}$ state, $s$, and to the $D_{3/2}$ state, $t$,
\begin{equation}
    \text{P}_{D_{5/2}} = \frac{s}{s+t}\text{P}_{D}
\end{equation}
which gives the fitting function
\begin{equation}
     \text{P}_{D_{5/2}} = \frac{s}{s+t}\left[1-\text{exp} \left(\frac{-a\Gamma}{(\omega-\omega_0)^2+\Gamma^2} \right)+c\right],
\end{equation}
where $a$, $\Gamma$, $\omega$, and $c$ are fitting parameters and the $s$ and $t$ branching fractions are from \cite{Fan2019a}.

\section*{Appendix B: Radium Ion Frequency Summary}\label{appendix:sum}

To verify that the measurements are self-consistent we perform an energy-level optimization with the reported $S_{1/2}\rightarrow P_{3/2}$ transition frequency and other Ra$^+$ transition frequencies reported in \cite{Fan2019, Holliman2019} (see Table \ref{table:frequency_summary}). If we fix the ground state at zero energy, there are seven transitions between four excited states which give three degrees of freedom. We use the energy-level optimization code package LOPT \cite{KRAMIDA2011419} to find the energy values that best agree with our reported frequencies. We calculate the residual sum of squares to be 0.14, which, being less than one, indicates that the measured frequencies are self-consistent. 

\begin{table}[ht] \centering
\begin{threeparttable}
\caption{Summary of $^{226}$Ra$^{+}$ transition frequency measurements. All units are GHz.
\label{table:frequency_summary}}
\begin{ruledtabular}
\begin{tabular}{lcccc}
    Transition & This work and \cite{Fan2019, Holliman2019}  \\ 
    \midrule
    $S_{1/2}\rightarrow$ $P_{3/2}$ & \SI{785722.11\pm0.03}{} \\
    $S_{1/2}\rightarrow$ $P_{1/2}$ & \SI{640096.63\pm0.06}{} \\
    $D_{3/2}\rightarrow$ $P_{3/2}$ & \SI{423444.39\pm0.03}{} \\
    $S_{1/2}\rightarrow$ $D_{5/2}$ & \SI{412007.701\pm0.018}{} \\
    $D_{5/2}\rightarrow$ $P_{3/2}$ & \SI{373714.40\pm0.02}{} \\
    $S_{1/2}\rightarrow$ $D_{3/2}$ & \SI{362277.68\pm0.05}{} \\
    $D_{3/2}\rightarrow$ $P_{1/2}$ & \SI{277818.95\pm0.08}{}\tnote{a} \\
\end{tabular}
%\vspace*{-\baselineskip}
\end{ruledtabular}
\begin{tablenotes}
\small
\item [a]Frequencies calculated from a sum of direct measurements in $^{226}$Ra$^{+}$.
\end{tablenotes}
\end{threeparttable}
\end{table}

%\bibliography{radium_five}


\end{document}